# As Transparent as Possible, System Implementation Governance Using Process Mining

Negin Maddah[1]

Department of Mechanical and Industrial Engineering, Northeastern University, Boston, MA, 02210

## Acknowledgments

Gratitude is extended to Celonis Academy for providing access to their advanced software, significantly enhancing research in the field of process mining.

## Data Availability Statement

The data that support the findings of this study are available from the corresponding author upon request.

## Abstract

This paper advocates for guiding an effective system implementation approach at a business process level. It details a case study of a food product manufacturer that transitioned to a new local information system. 41 units' data (10160 cases) over the pre-maturity phase of the system were then structured into event logs and analyzed. This analysis identified deviant process paths, questioning whether the new system efficiently supports procurement operations immediately post-implementation. The findings reveal critical implementation risks with conformance-checking of the as-is process with the to-be process model; this includes incomplete cases, unauthorized activities, irregular payment practices, stemming from organizational bottlenecks, or violating internal control regulations. Challenges are attributed to technical shortcomings in system design and cultural misalignments, necessitating immediate interventions or longer-term cultural and training solutions. This study's contribution is its demonstration of a transparent, process-driven approach to system governance, highlighting the strategic benefits of this integration for organizational management.

## Introduction

The dynamic landscape of organizational management is continually reshaped by technological advancements, particularly through the implementation of information systems designed to integrate various organizational functions into a singular, coherent framework. Despite their potential to transform business processes, their deployment is fraught with challenges [1], often characterized by high implementation failure rates and the daunting task of aligning system functionalities with complex business needs. In this context, the emergent field of process mining can offer a granular view of how business processes unfold within an organization [2], providing insights into the actual versus the ideal functioning of processes as envisaged during the system's design phase.

Central to the utility of process mining is its ability to conduct conformance-checking—a methodological approach that scrutinizes the alignment between the executed (as-is) processes and the planned (to-be) process models [3]. This capability is particularly critical in the post-implementation phase of systems, where identifying discrepancies and deviations can lead to targeted interventions, thereby enhancing process efficiency and regulatory compliance.

[1] Corresponding Author: Negin Maddah (PhD Candidate at Northeastern University). maddah.n@northeastern.edu

This study seeks to bridge this gap by employing process mining techniques, with a focus on conformance-checking, to oversee the system migration process within a food industry company. By analyzing a dataset encompassing 41 business units over eight months, this research aims to shed light on the practical benefits and challenges of leveraging process mining for system implementation and governance. The objective is to demonstrate how process mining can provide a more agile, transparent, and effective framework for system migration, thereby contributing to the broader discourse on enhancing organizational governance through system interventions or employee management [4].

The structure of this paper is organized to facilitate a comprehensive understanding of the study’s approach and findings. After presenting an overview of the related topics in the literature, the paper will detail the methodology designed for this specific application, followed by the implementation of this methodology on real-world data. This includes visualizations and analyses that provide insight into the practical implications of process mining for system migration governance. The concluding section of the paper will offer reflections on the study’s outcomes, and directions for future research. Through this structured exploration, the paper aims to contribute significantly to the realm of system governance and implementation, offering a new perspective that integrates process mining with organizational strategy and operational efficiency.

## Literature Review

The advancement of process mining has revolutionized Business Process Management (BPM) [5], introducing a novel lens through which organizations can examine the lifecycle of their business processes. As a bridging mechanism, process mining fills the void between traditional model-based process analysis techniques, such as BPM strategies, and data-centric analytical approaches, including machine learning and data mining [6]. Van der Aalst has underscored and subsequent studies across various sectors have validated that process mining offers an in-depth insight into actual business operations by tapping into the data embedded within information systems [7]. Fundamentally, process mining facilitates process discovery, conformance-checking, analysis of process variants, and assessment of process performance [8], [9]. While this study touches upon all these facets, it specifically leans towards the conformance-checking method for system governance, which aims at pinpointing and quantifying the deviations between the model and the actual process logs [3].

Documenting and storing the conceptual models resulting from BPM methodologies poses a challenge for process mining applications. Introduction of the Process Repository by Celonis an integrated suite of tools that not only improves process documentation with standard Business Process Model and Notation (BPMN) [10] but also integrates seamlessly with process mining applications [11]. The Process Repository within the Celonis Intelligent Business Cloud acts as a hub for documenting and managing business processes, analyzing process performance and conformance, and enabling integration with the BPM tool [11]. This feature bridges the gap between the actual executed (as-is) processes and the ideal (to-be) models, simplifying the conformance-checking process for practitioners and researchers in process mining.

The study's conformance-checking between the to-be process model and the as-is process elucidates the tangible benefits of recent advancements in this domain. Conformance-checking of a newly implemented information system serves a dual purpose: it not only verifies the model's accuracy in reflecting the system’s design to facilitate process optimization but also identifies deviations and potential violations of internal control regulations [12], [13]. Although process mining has benefited research and industry for anomaly detection purposes recently [14], the proposed approach of this paper signifies a move towards more agile, transparent, and effective governance of system migration practices.

The adoption of Enterprise Resource Planning (ERP) systems represents a pivotal development in organizational management, offering an integrated solution to manage various functions within a unified framework. ERP systems are described as comprehensive software packages that incorporate different organizational functions into a single database with multiple modules [15]. These information systems are designed to enhance operational efficiency and strategic insight by ensuring smooth process integration and data accessibility across the organization [16]. Software giants like SAP are at the forefront of ERP deployment, continuously updating their offerings to enable organizations to make swift decisions on real-time issues, thereby ensuring day-to-day business process control [17]. However, despite the transformative promise of ERP systems, their implementation and the full realization of their benefits come with significant challenges, including high failure rates. These challenges highlight the critical need for thorough post-implementation reviews [1]. Traditional methodologies for post-implementation review tend to bypass detailed business process-level analysis in favor of broader metrics such as Return on Investment (ROI), which may not fully capture the impact of system implementations on an organization's operational dynamics [18].

The introduction of integrated systems, despite the challenges in adoption across various departments, has the potential to significantly enhance organizational performance management [19]. In an extension of process mining's capabilities, [20] introduce a holistic framework aimed at assessing the performance of business units based on the data of their integrated information systems, enabling managers to make more informed decisions, thereby facilitating the effective prioritization of resources for overall improvement. Expanding the versatility of process mining across other sectors, [21], [22], and [23] have demonstrated its effectiveness in healthcare [24], e-commerce, and sales systems, respectively. This showcases integrated systems' utility in real-time auditing [25], reengineering of business processes, and addressing operational inefficiencies across various industries [26].

However, there exists unexplored areas in the governance of pre-mature information systems, which can be addressed by process mining, an application has been underrated in the literature. The insights from [15] and [16] implicitly underscore the critical role that process mining can have in system migration governance. This perspective not only can promote transparency in the evaluation of ERP implementations but also ensures that system functionalities are closely aligned with organizational objectives. Through the lens of process mining, organizations can navigate the complexities of system migration with greater precision, ensuring the alignment of new or evolved information systems with the dynamic requirements of the business landscape.

[27] expand on the utility of process mining, specifically in the context of adaptive systems, facilitating a deep understanding of the evolutionary nature of business processes post-ERP implementation, which allows for the identification of the driving forces behind changes. [28] advocate for the application of process mining techniques in conducting ERP post-implementation reviews at the business process level. Their investigation within the procurement department of an agricultural chemicals company revealed both adherence to and deviations from standard procedures, attributable to factors like reliance on hardcopy documentation and urgent procurement needs. These deviations led to process inefficiencies, underscoring the importance of granular, process-level examination to uncover and address implementation shortcomings.

This study underscores the benefits of process mining in the context of system implementation within a food industry enterprise. By leveraging operational data over the pre-mature phase of a system implementation, the paper addresses a tangible system migration challenge. It aims to enrich the discourse on process mining's role in system governance, particularly through a conformance-checking lens. The proposed methodology illuminates the application of process mining at an organizational level [3], highlighting its impact on operations and value generation within entities.

## Methodology

To perform a comprehensive process mining analysis, it's imperative to collect log data detailing the sequence of activities within the integrated system. This data must include an identifier to trace each case through its process journey. Initially, the business process is standardized and modeled based on system requirements analyses, a foundational step typically completed in the system design phase. Accordingly, BPM interviews may only be necessary in particular scenarios to further refine this model. For the process mining analysis to commence, extracting three critical components from the database is essential: the case ID, the activity, and the timestamp for each activity. These elements are crucial for tracking and analyzing the workflow within the system.

As illustrated in Figure 1, the process begins with modeling the to-be process, which represents the ideal workflow envisaged during the system's design phase. Following the system's introduction to the organization, an event log capturing the real process's execution is extracted for analysis. This real-time activity data, spanning various business departments and roles, is stored within the integrated system. Process mining techniques then extract the as-is process from this dataset. Such a process map not only clarifies the workflow dynamics within the organization but also establishes a benchmark for comparison against the to-be model.

Employing conformance-checking aligns the real process data with the to-be model. This step generates detailed statistics on conforming and non-conforming cases, facilitating a root-cause analysis. This analysis spotlights discrepancies or violations in the real process execution [29], introducing the option to create an allowlist for deviations considered acceptable under specific conditions. The primary emphasis, however, is on identifying and remedying the non-conforming cases. This comparison using predefined key performance indicators (KPIs)—whether time-based or other process-specific metrics—enables the assessment of the system's implementation efficiency and compliance. Celonis Process Query Language (PQL) [30] enables defining any process-oriented KPI in its software. This is instrumental in identifying areas of non-conformance and understanding these deviations' implications on the organization's operational efficiency and strategic objectives. Based on the findings, improvement strategies—ranging from short-term immediate corrections to long-term cultural or procedural adjustments—can be integrated into the operational system for enhanced performance and alignment with organizational goals.

Upon identifying non-conforming cases through conformance-checking, it's pivotal to categorize the violations based on their root causes, which typically fall into two primary groups: system design issues and human-related issues. This categorization pinpoints the source of each discrepancy and formulating appropriate corrective measures. In each phase of the analysis, it is possible to filter and target specific cases associated with a particular type of violation. This targeted approach allows for a detailed examination of the time, case ID, department, activity, product, or any other pertinent details related to these cases. This nuanced analysis enables a more informed decision-making process, allowing stakeholders to understand whether the cause of a violation stems from inadequacies in the system's design or from human error, such as misuse or misunderstanding of the system's functionalities. By distinguishing between these cause categories, the organization can tailor its improvement strategies more effectively. For system design-related issues, the focus might be on technical enhancements or modifications to the system itself. Conversely, for human-related violations, the solution may involve targeted training programs, workflow adjustments, or policy changes to address and mitigate the identified problems.

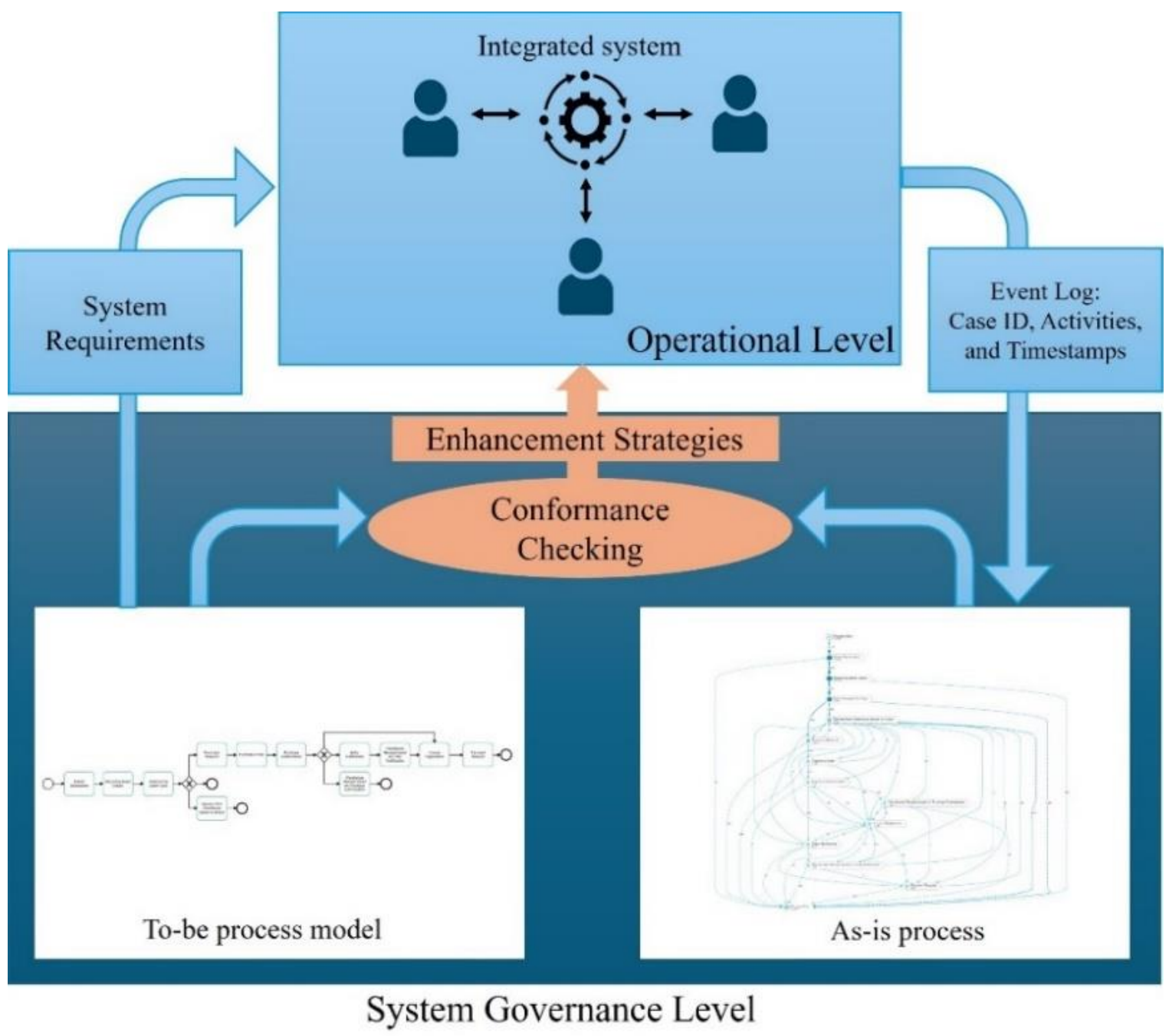

*Figure 1. Research Methodology Overview*

## Implementation and Results

This section presents the application of the outlined methodology for overseeing the implementation of a new local ERP system in a food industry company. By leveraging the system's data—initially considered to be of limited utility—collected during the first eight months following its launch, we aim to guide the organization in prioritizing system improvement strategies utilizing four critical process mining lenses: process overview, process explorer, social mining, and conformance-checking.

This analysis was structured around key lenses to thoroughly investigate and interpret the data: (1) Data Overview: Providing insights into the scope of the data and system analyzed. (2) Process Overview: This lens provides a macro view of the procurement process, identifying the main paths taken by the majority of cases. It helps in understanding the general flow of activities and detecting any unexpected or irregular paths. (3) Process Explorer: Offering a micro-level analysis, the Process Explorer dives into specific cases, enabling the examination of detailed sequences of events. This lens is crucial for identifying deviations and exploring the variability within process paths. (4) Social Mining: This aspect of the analysis focuses on the interaction between different roles within the organization. By examining the social aspects of the process, it's possible to identify collaboration patterns, bottlenecks caused by workload distribution, and potential communication gaps. (5) Conformance-checking: The core of our methodology, conformance-checking, assesses the alignment between the actual process execution and the designed process model. This lens is instrumental in pinpointing discrepancies, assessing the severity of deviations, and identifying non-conforming cases for further root-cause analysis.

The ultimate objective of applying these process mining lenses is to identify and prioritize system improvement strategies within the organization. By systematically analyzing the procurement process from multiple perspectives, we can uncover actionable insights into how the system can be optimized. The findings from each lens contribute to a comprehensive understanding of both the strengths and weaknesses of the current ERP system implementation. This holistic view facilitates informed decision-making regarding where to allocate resources and efforts to achieve the most significant impact on system performance and organizational efficiency.

In the following sections, the findings are presented from each of these lenses and discuss the implications for system improvement and organizational strategy.

## *Data Overview*

Understanding the procurement process is crucial, as it constitutes a fundamental component of organizational operations across various sectors [31]. This study is focused on the procurement process, aiming to shed light on the universal challenges that businesses encounter. To tackle the issue of diverse process paths, we conducted an analysis of a wide array of requests—10,160 cases—spanning 41 business units within the food production company in focus. This dataset covers system interactions by employees from April 2022 to December 2022. It's worth noting that while the pre-processing of system data into an event log represents a critical phase in process mining analysis [32], [33], this aspect falls outside the scope of our current investigation. An initial summary of the analyzed data is presented in Table 1.

*Table 1. Data Scope of the Study*

| Data Scope Element | Total amount/range |
|---|---|
| cases | 10160 |
| events | 42637 |
| timestamp | 2022-04-05 - 2022-12-03 |
| Units | 41 |
| Units' categories | 10 |
| Requests categories | 10 |

Figure 2 presents a detailed BPMN diagram illustrating the sequence and relationship of the procurement activities within the company's newly implemented ERP system. This diagram serves as the benchmark for the conformance-checking analysis detailed later in the paper. The activities reflect the ideal workflow and decision points within the procurement process. Here's a breakdown of the activities as visualized: (1) Needs Declaration: The process initiates with the declaration of procurement needs by the relevant department or team within the organization. (2) Recording Need Details: Following the initial declaration, detailed specifications and requirements of the needed goods or services are recorded. (3) Determining Action Type: At this juncture, based on the detailed needs, a decision is made—either to fulfill the requirement directly from the warehouse or to initiate an external purchase request. (4) Delivery from Warehouse based on Action: If the decision is to source from the warehouse, this phase triggers the process for the initiation of delivery to the requesting department. (5) Purchase Request: For needs that cannot be fulfilled by the warehouse, a purchase request is initiated to procure the required goods or services externally. (6) Purchase Order: A purchase order is generated. (7) Purchase Authorization: Procurement activities require authorization post-order placement, documented as the purchase authorization activity to finalize the order. (8) Entry Notification: Once the vendor dispatches the order, an entry notification is created to alert the receiving department. (9) Warehouse Receipt based on Entry Notification: Goods received are documented and stored in the warehouse based on the entry notification. (10) Invoice Registration: the invoice registration activity ensures that all financial documentation is accurately recorded. (11) Payment Request: A payment request is made to the finance department. (12) Warehouse Receipt based on Purchase Authorization: Any additional goods received requiring post-authorization are documented and processed into the warehouse.

This BPMN diagram (Figure 2) not only delineates the process flow but also highlights key decision points and transitions between activities, making it an invaluable tool for both understanding the procurement process and conducting the conformance-checking analysis.

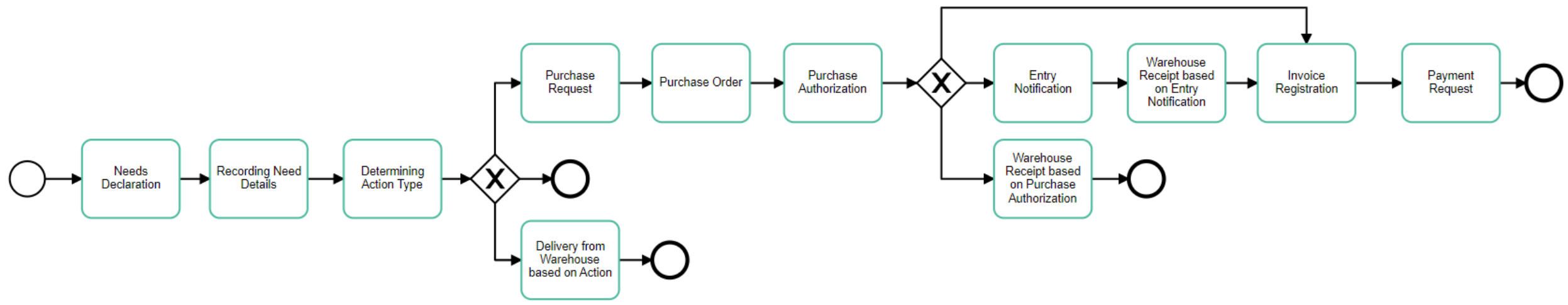

*Figure 2. To-be Business Process Model*

To illustrate the diversity of the data concerning the requests and their corresponding units, Figure 3 provides a representation of the volume of requests per product and by unit category. A logarithmic scale was adopted due to the significant imbalance observed in the distribution of requests across the units, with a notably higher concentration in two specific categories: HR and Finance, and Project and Infrastructure Development. Given the analysis encompasses 41 distinct business units, a decision was made to consolidate these units into broader categories based on their functional similarities, enhancing the clarity and interpretability of the presentation.

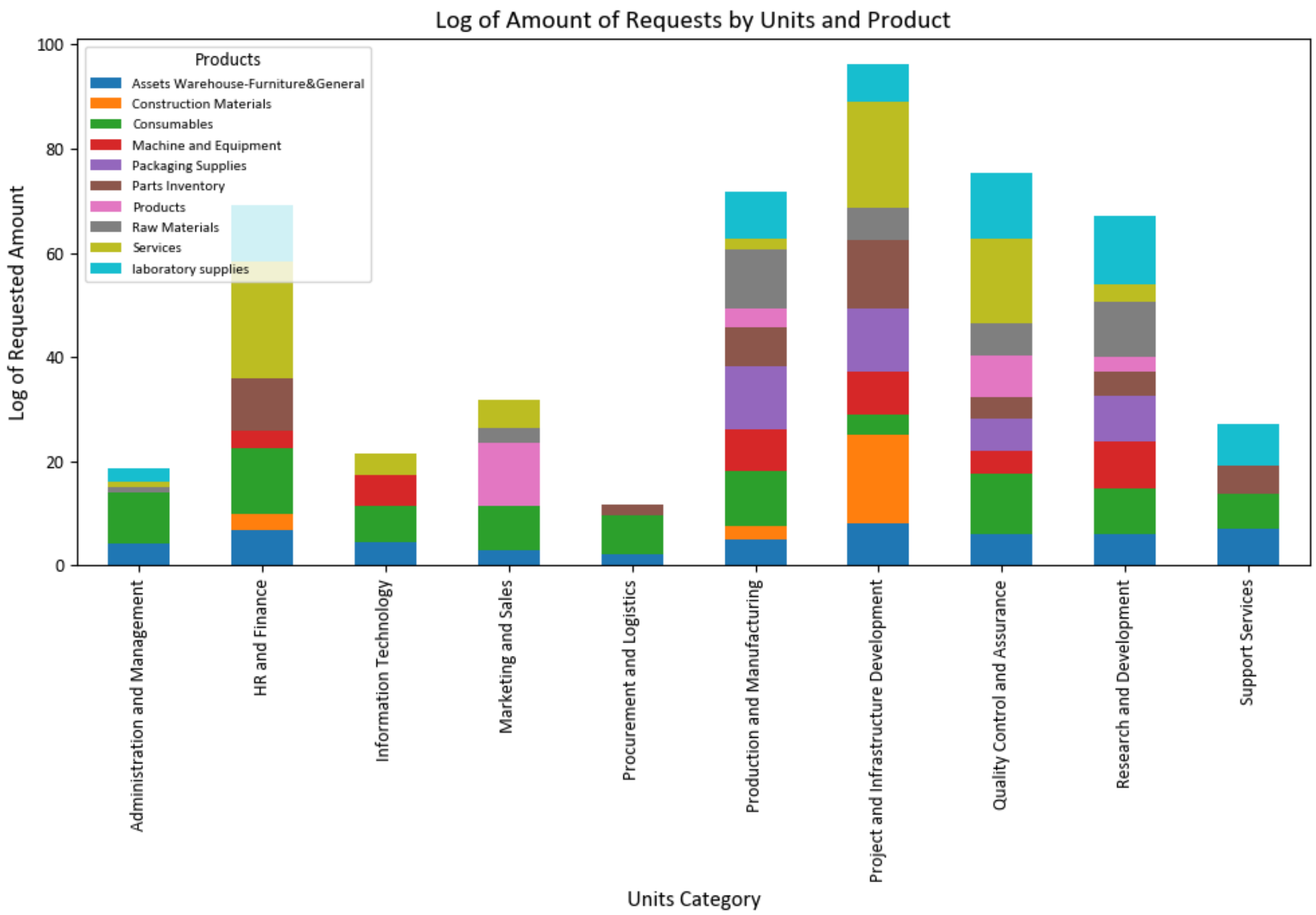

*Figure 3. Variety of the Organizational Units Categories and Their Requests in the Study*

## *Process Overview*

In assessing the dynamics of the procurement process within the selected time frame, several key metrics have been analyzed. On average, there were 48 cases per day, translating into approximately 201 events daily. Notably, the average duration for a case to progress from initiation to conclusion across the system was observed to be 14 days.

The evolution of these metrics throughout the study period is depicted in Figure 4, utilizing Celonis business intelligence software for data analysis and visualization.

A significant observation from this analysis is the increased system usage as the study period progressed. This trend was anticipated and reflects a growing familiarity and integration of the new system within the organization’s daily operations. However, more critical than the sheer increase in system usage is the noticeable improvement in throughput time—the interval from the start to the end of the process. This metric's enhancement indicates that processes are being completed more swiftly in the latter months of the system's implementation compared to its initial phase.

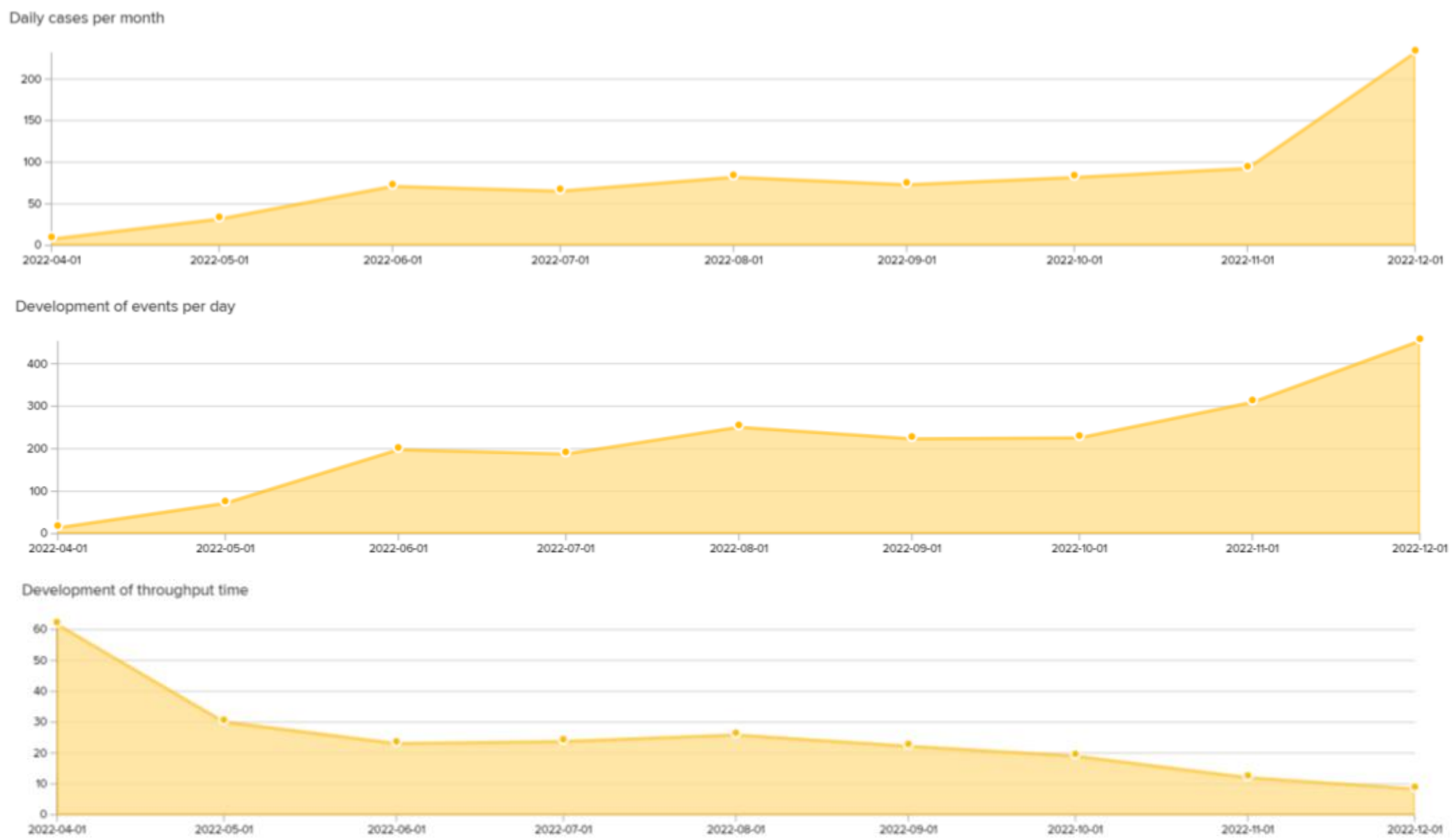

*Figure 4. The Trends of Cases, Event, and the Duration of the Process*

The distribution of the average throughput time for the entire process is depicted in Figure 5, revealing a median completion time of 6 days for cases. However, a significant number of cases exhibit exceptionally long completion times, which are not only unreasonable but also indicative of potential inefficiencies within the process. These outliers warrant a detailed investigation, including an examination of the specific requests, the departments and employees involved, and any discernible patterns that may be contributing to delays. Particularly concerning are the cases represented in the middle bars of the histogram (since the final bars include the incomplete cases), which concluded within approximately 80 days. These durations, considerably longer than the median, suggest specific types of requests or services might inherently require more time, necessitating a deeper dive into the nature of these requests for targeted improvement measures. Notably, the robustness of these findings persists even after accounting for inherently time-intensive requests by filtering out these requests.

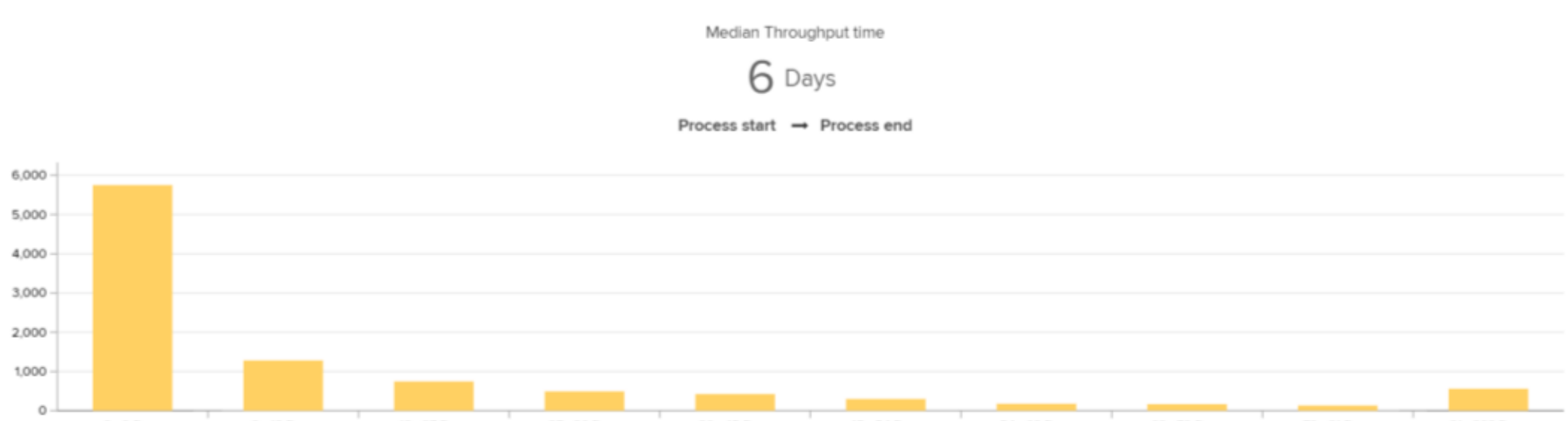

*Figure 5. The Histogram of the Process’s Duration*

Given these insights, a more granular, process-oriented analysis is imperative to accurately pinpoint the bottlenecks within the procurement process. Utilizing process mining techniques, as shown in Table 2, we have identified five critical bottlenecks that significantly impact the process flow. Furthermore, Figure 6 highlights the uneven distribution of activities throughout the process, further illustrating the areas where inefficiencies concentrate.

*Table 2. Identified Bottlenecks of the Process*

| | Source Activity | Destination Activity | Throughput time (workdays) | %Cases affected |
|---|---|---|---|---|
| 1 | Purchase request | Purchase order | 28 | 17% |
| 2 | Determining action type | Delivery from warehouse | 13 | 35% |
| 3 | Recording needs details | Determining action type | 6 | 89% |
| 4 | Determining action type | Purchase request | 1 | 35% |
| 5 | Needs declaration | Recording need details | <1 | 99% |

These bottlenecks, ranging from less than a day to 28 days in throughput time, affect a significant portion of the cases, underscoring the necessity for targeted interventions to streamline these specific transitions within the process. The largest bottleneck appears to occur between the stages of "Purchase Request" and "Purchase Order," impacting 17% of cases with a notable delay of 28 workdays, suggesting inefficiencies in the procurement authorization or vendor negotiation phases.

Figure 6 further visualizes the disparity in activity volumes throughout the process, offering a visual representation that complements the bottleneck analysis provided in Table 2. This combination of quantitative and visual analysis forms a robust foundation for developing effective strategies to address the identified inefficiencies, with the ultimate goal of reducing throughput times and improving overall process efficiency.

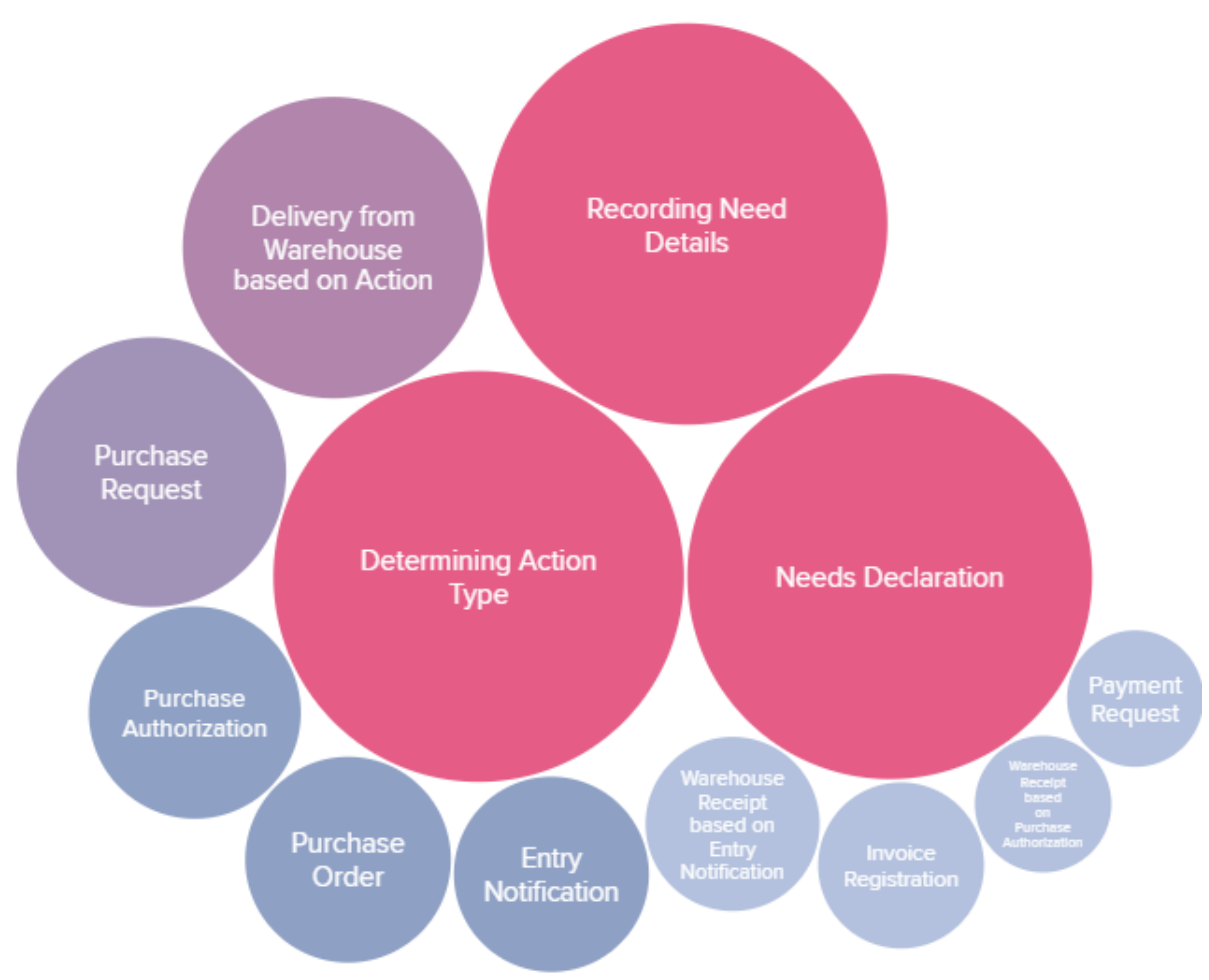

*Figure 6. Activities Occurrence Volume*

## ***Process Explorer***

Figure 7 offers a revealing glance at the actual dynamics of the procurement and request fulfillment process as employees navigate the newly implemented system. This visualization uncovers a multitude of unexpected pathways, numerous instances of unmet requests, and regulatory shortcuts—all of which become apparent upon initial observation. Such intricacies highlight the complexity and challenges inherent in adopting a new system, presenting a real-world picture of the process flow across various business units over an extended period.

This form of visualization stands out as exceptionally insightful and transparent, providing a direct view into the operational nuances of a significant number of units. It illustrates not only the anticipated paths but also the unforeseen routes that cases might follow, shedding light on areas of inefficiency and non-compliance that might not be as easily detected through traditional analysis methods. The numbers on each arc of this process map show the cases volume following them.

The insights gained from this Process Explorer view reinforce the belief that system migration governance is a critically underappreciated application of process mining. The ability to offer real-time feedback on the system's functioning—bypassing the need for labor-intensive interviews and minimizing the risk of human error—underscores the value of process mining in guiding and refining the implementation of new systems. This approach enables organizations to quickly identify and address issues, ensuring smoother transitions and more effective system integration.

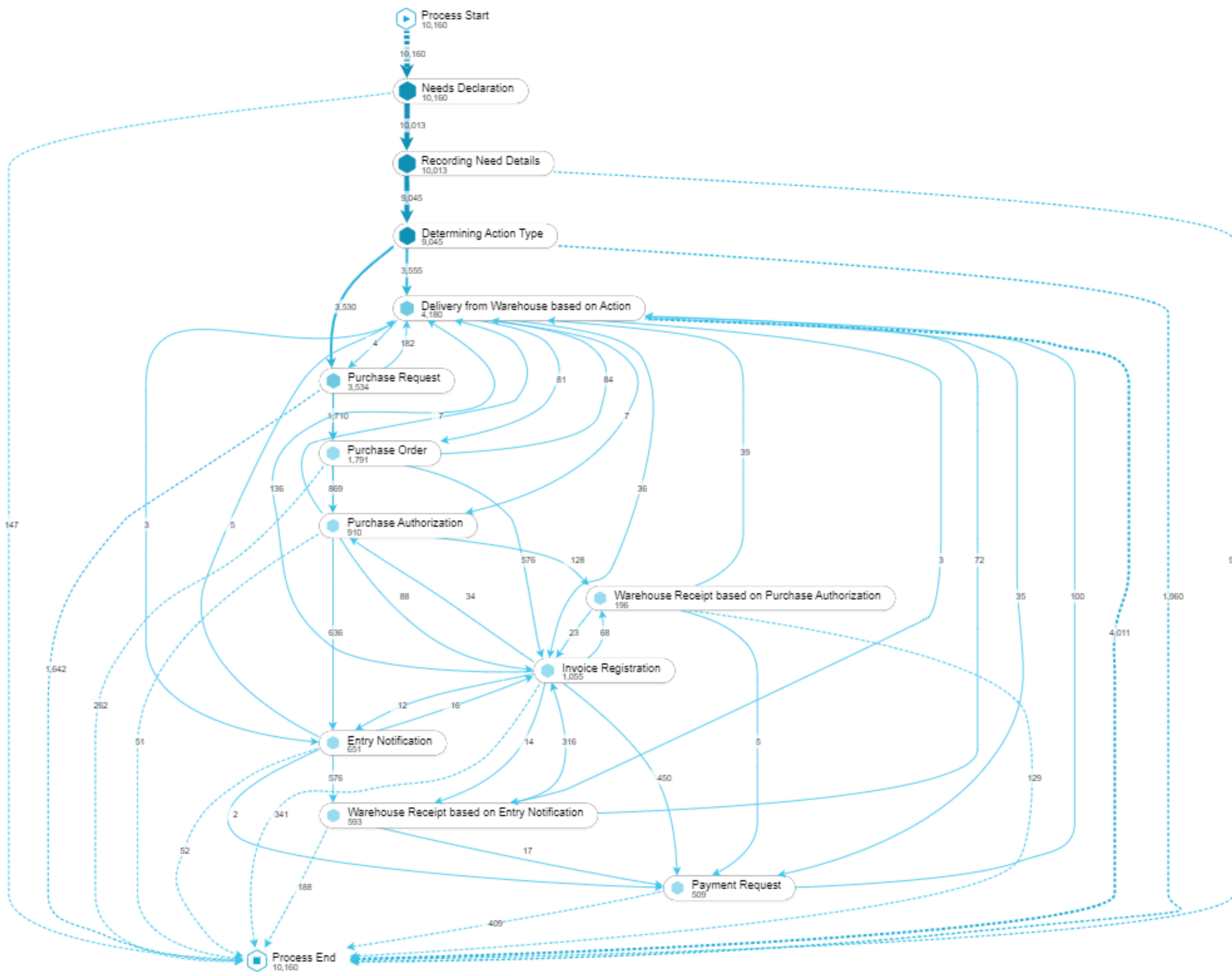

*Figure 7. Process Map, Showing What is Happening in Reality*

## ***Organizational Social Miner***

Through the lens of social mining, an analysis of the organization's departments reveals that on average, 11 departments categories executed an activity per day, with an average daily event count of 17 per category, and an average caseload of 248 cases per category. This data highlights the increasing engagement of employees with the system over the analyzed period, indicating a growing comfort and proficiency among staff in navigating the new system's functionalities. The upward trend in system utilization is detailed in Figure 8, illustrating how adoption has evolved during the study period.

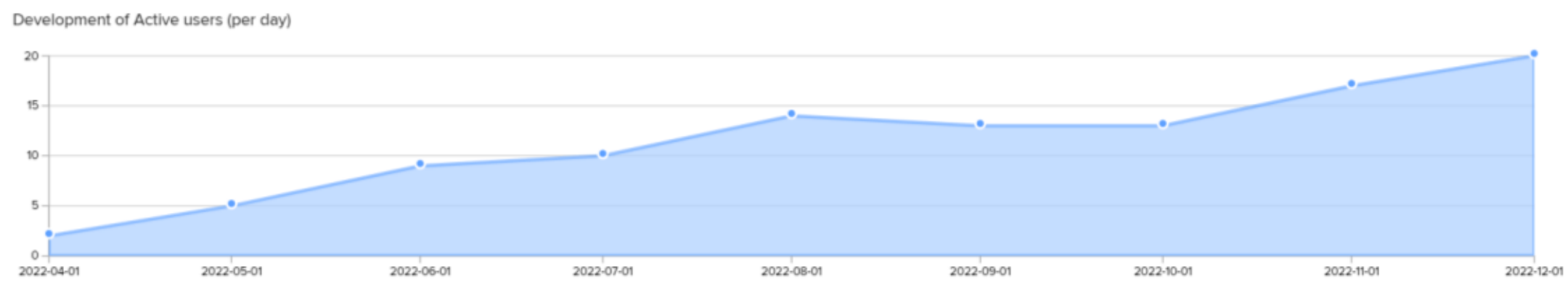

*Figure 8. Development of Active Business Units over Time*

The variability in engagement levels among departments underscores the opportunity for tailored educational interventions to expedite system adoption across the business. By strategically deploying training initiatives, organizations can ensure a quicker alignment of departmental activities with the overall goals of system migration, facilitating a smoother transition process.

The impact of different departmental goals and workload distributions on system usage rates becomes apparent when considering the broader activity landscape provided by Figure 9. This visualization reveals the uneven collaboration on procurement activities across business units, reflecting the natural variance in responsibilities and job functions. Despite this imbalance, such insights are invaluable for orchestrating a comprehensive approach to social management within the system migration framework.

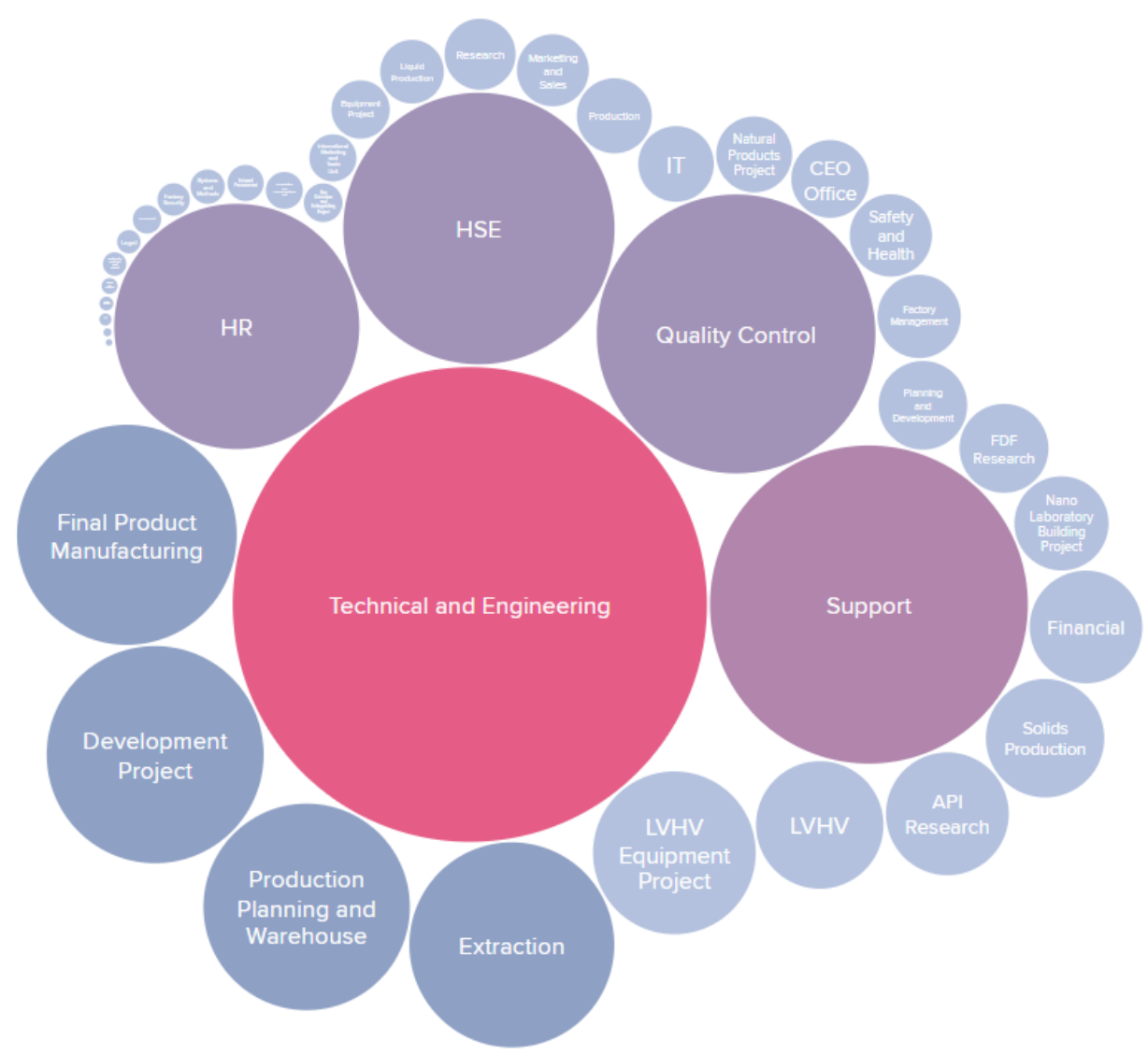

*Figure 9. The Volume of Business Units System Deployment*

## ***Conformance-Checking***

The conformance-checking stage in process mining plays a pivotal role in comparing the as-is process—how the process actually unfolds pre-maturity phase—with the to-be model, which represents the ideal process flow as envisioned during the system's design phase. Despite an organization's clear vision of its desired business processes, the actual post-implementation scenario often deviates from this ideal. Utilizing a BPMN version of the to-be model facilitates an in-depth analysis of deviations occurring with the use of the new system within the organization.

From the analysis, it's observed that only 55% of the cases conform to the process variations introduced in the to-be model. In numerical terms, this translates to 5,600 conforming cases out of a total of 10,160, leaving 4,560 cases as non-conforming. These groups can be further dissected and examined through process overview, exploration, and social mining techniques. The historical trend of conformity, depicted in Figure 10, indicates that no significant regulatory improvements have been applied during the first eight months following the system's introduction. Notably, a portion of the non-conforming cases in the latter months can be attributed to incomplete cases, which, when excluded from the analysis, alters the conformity landscape.

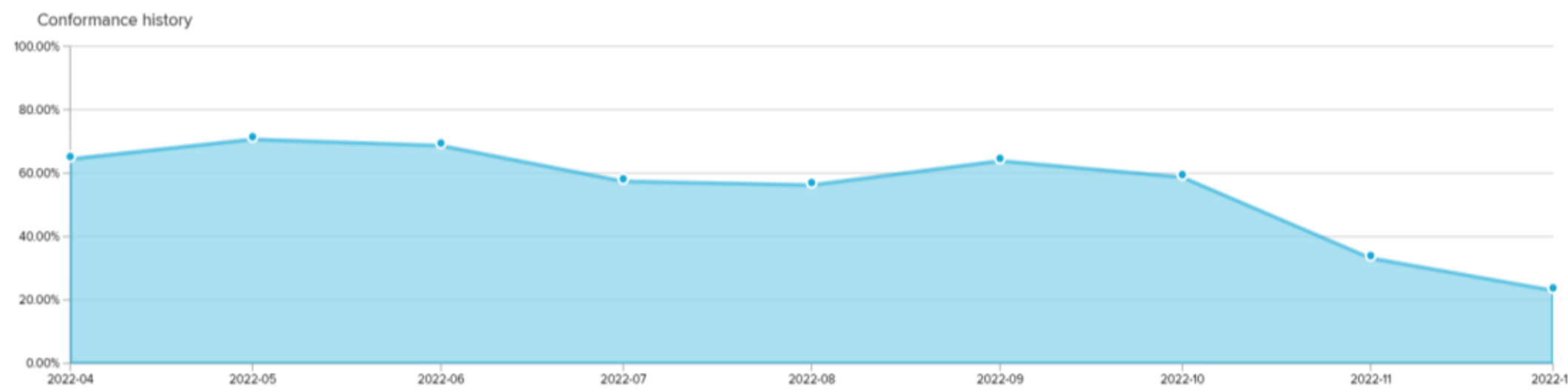

*Figure 10. Conforming Cases Trend over Time*

Upon excluding the incomplete cases for a more focused exploration, the distribution of conforming versus violating cases shifts to 8,850 versus 1,130. This adjustment reveals a pattern of initial increase in process violations, which subsequently decline, indicating a gradual improvement in process adherence. This trend suggests that while enhancements are evident, they could potentially have been realized earlier with the full transparency afforded by process mining. For the remainder of the analysis, incomplete cases are set aside. Although these cases warrant separate investigations as potential violations due to unfulfilled requests, including them could compromise the robustness of the study's outcomes. This is primarily because cases initiated towards the end of the study's timeline may not reasonably be expected to conclude by December 2022.

Defining various KPIs for contrasting conforming and violating cases is essential. These KPIs should be crafted to reflect both operational strategies and internal control metrics, allowing organizations to align the performance evaluation of the new system with specific objectives, whether they be time, action, or product-related. Identifying the root causes of internal controls discrepancies enables the refinement of the process toward a model free from regulatory issues, thereby optimizing the system's overall efficiency and compliance.

The conformance-checking method offers insightful revelations into the operational dynamics of a newly implemented system, particularly through a timely analysis of process adherence. This analysis demonstrates that violating cases, on average, take 38.7 days longer to complete than conforming cases, which have an average throughput time of 13.4 days. This significant disparity not only highlights the deviation from the desired process path but also underscores the inefficiency in process execution, with violating cases moving at a slower pace. Moreover, the analysis reveals that violating cases typically involve an increased number of steps, averaging 4.1 more steps than conforming cases. This observation aligns with the previous finding, as the additional steps contribute to the extended throughput time, further deviating from the optimal process flow defined in the to-be model.

The study's conclusive findings, presented in Table 3, identify specific violations through conformance-checking. A prevalent issue in newly implemented systems is the incorrect order and sequence of process steps, which contravenes internal control regulations. A notable example of this violation is the placement of orders in the system post-invoice registration without prior authorization from an independent authority—a practice observed in 576+88 cases (violations 1 and 6 in Table 3). Running a root-cause analysis for these two violations revealed that 192 are correlated with some specific business units and projects. Another significant category of violations involves activities related to warehouse deliveries, particularly when these are followed by purchase requests or payment activities, suggesting a misalignment with inventory management practices. These violations necessitate a thorough investigation into the system governance level to pinpoint the origins of these discrepancies (violations 3, 4, 5, 7, and 8 in Table 3). Additionally, certain violations are influenced by overarching management strategies and decision-making processes, such as the sequence of registering an invoice before documenting warehouse receipts or executing warehouse deliveries (violations 2 and 9 in Table 3). These instances may reflect higher-level strategic decisions but nonetheless represent deviations from the established process model.

*Table 3. Detected Process Violations based on Conformance-Checking Approach*

| Violation | Source activity | Destination activity | cases affected | Effect on throughput time (days) | Effect on steps per case |
|---|---|---|---|---|---|
| 1 | Invoice registration | Purchase order | 6% (576) | +39 | +3.4 |
| 2 | Invoice registration | Warehouse receipt | 3% (316) | +43 | +6.1 |
| 3 | Delivery from warehouse | Purchase request | 2% (182) | +42 | +3.1 |
| 4 | Delivery from warehouse | Invoice registration | 1% (136) | +44 | +5.2 |
| 5 | Delivery from warehouse | Payment request | 1% (100) | +47 | +5.1 |
| 6 | Invoice registration | Purchase authorization | 1% (88) | +29 | 4.9 |
| 7 | Delivery from warehouse | Purchase order | 1% (84) | +35 | +3 |
| 8 | Delivery from warehouse | Warehouse receipt based on purchase authorization | <1% (39) | +16 | +5.2 |
| 9 | Invoice registration | Delivery from warehouse | <1% (36) | +20 | +4.8 |

This analysis underscores the utility of conformance-checking in illuminating areas of process non-adherence, inefficiency, and regulatory misalignment. By identifying and understanding the nature and causes of these violations, organizations can implement targeted interventions to rectify these issues, thereby enhancing the effectiveness and compliance of the system implementation.

## Conclusion

This study showcases the complexities of information system implementation within a food industry company, employing process mining techniques to uncover the gap between the idealized process models and the actual post-implementation realities. Through the analysis of data from 41 business units over an eight-month pre-maturity phase, significant deviations, inefficiencies, and non-compliances were identified within the newly integrated system. The findings highlight the critical role of process mining, especially conformance-checking, in detecting operational bottlenecks and compliance issues, such as unauthorized transactions and errors in process step sequences causing internal auditing issues.

These insights reveal crucial areas for improvement in system design, user training, and internal controls to streamline process flow and ensure regulatory compliance. Strategic recommendations include targeted training to mitigate common errors, system redesigns to prevent unauthorized practices, revision of internal controls for stronger compliance, and a continuous improvement approach anchored in root-cause analysis.

Nonetheless, the study's scope, confined to a single organization and a specific ERP system, along with the potential oversight of process variation and organizational culture impacts, suggests areas for future exploration. Further research should expand the scope to multiple industries, integrating causal analysis for deeper insights into root causes. Furthermore, a more comprehensive analysis needs to be done before and after the implementation of such a framework including both pre-maturity and mature phases of the information system. Such work is essential to understand system migration dynamics, overcome the current study’s limitations, and enhance the generalizability of its findings.